\documentclass{emulateapj}

\shorttitle{Black Hole Binary}
\shortauthors{Makino and Funato}

\newcommand{\msun}{M_{\odot}}
\def\apgt{\ {\raise-.5ex\hbox{$\buildrel>\over\sim$}}\ }
\def\aplt{\ {\raise-.5ex\hbox{$\buildrel<\over\sim$}}\ }

\begin{document}


\title{Evolution of Massive Black Hole Binaries}


 \author{Junichiro Makino \altaffilmark{1}
          and
         Yoko Funato\altaffilmark{2}}
 \altaffiltext{1}{Department of Astronomy, University of Tokyo, 7-3-1 Hongo,
         Bunkyo-ku,Tokyo 113-0033, Japan}
 \altaffiltext{2}{Department of General System Studies, University of Tokyo,
         3-8-1 Komaba, Meguro-ku, Tokyo 153-8902, Japan}


\begin{abstract}

We present the results of large-scale $N$-body simulations of the
stellar-dynamical evolution of  massive black hole binaries at the
center of  spherical galaxies. We focus on the dependence of the
hardening rate on the relaxation timescale of the parent galaxy. A
simple theoretical argument predicts that a binary black hole creates
a ``loss cone'' around it. Once the stars in the loss cone are depleted, the
hardening rate is determined by the rate at which field stars diffuse
into the loss cone.  Therefore the hardening timescale becomes proportional to
the relaxation timescale. Recent $N$-body simulations, however, have
failed to confirm this theory and various explanations have been
proposed. By performing simulations with sufficiently large $N$ (up to
$10^6$) for sufficiently long time, we found that the hardening rate
does indeed depend on $N$. Our result is consistent with the simple theoretical
prediction that the hardening timescale is proportional to the
relaxation timescale. This dependence implies that  massive black
hole binaries are unlikely to merge within a Hubble time through
interaction with field stars and gravitational wave radiation alone.

\end{abstract}




\section{Introduction}

\subsection{Massive black hole binaries and the central
structure of ellipticals}

The possibility of the formation of massive black hole binaries in the
cores of  elliptical galaxies was first pointed out by
\citet[hereafter BBR]{Begelmanetal1980}. At that time, the possibility
of such events was purely theoretical, derived from two hypothesis. One is
that QSOs are driven by massive central black holes, with masses of the
order of $10^8 \msun$ or larger. The second is that most elliptical
galaxies, which might contain such massive black holes, are formed
through merging  of two galaxies. If both of the progenitor galaxies contain
massive black holes, these black holes sink to the center of the
merger product through dynamical friction and form a binary.

Though this scenario was purely theoretical at the time first proposed,
a considerable amount of observational support has been provided in the
last two decades.  First of all, there is now plenty of evidence
that many, if not most,  giant ellipticals contain massive central
black holes \citep{Magorrianetal1998}. Also, it has been suggested
that the black hole mass $M_{\rm BH}$ shows a tight correlation with the
spheroidal mass and the central velocity dispersion
\citep{Gebhardtetal2000,FerrareseMerritt2000}. The most
straightforward and natural explanation of the observed correlations
is that massive
galaxies are formed by merging of less massive galaxies, and that the
central black also grows through merging
\citep{KauffmannHaehnelt2000}.

This merging scenario has an additional important advantage: it
nicely explains the observed structure of the central region of massive
elliptical galaxies. In the 1970s and 1980s, observations revealed that
large ellipticals had large ``cores'', with a good linear
correlation between the size of the core and the size of the
galaxy. At that time, this correlation was thought to be a very strong
counter evidence against the merger hypothesis. The larger core size
of the larger galaxy implies that the central phase space density is
lower for larger galaxies. However, if larger galaxies had been formed
by collisionless merging of smaller galaxies, the central phase space
density would have been roughly conserved, and therefore the core size
would also have been roughly conserved.

Strictly speaking, the phase space argument only guarantees that the
central phase space density would not increase, and does not imply that it
is conserved. Nevertheless, numerical simulations of mergers
\citep{Faroukietal1983,White1979b,Okumuraetal1991} all demonstrated
that the core size would not increase significantly through merging.
And if some dissipational process implicit in the dynamics of gas clouds is
taken into account, most likely the central density would increase, rather
than decrease. Thus, it seemed very difficult to explain the observed
correlation between the core size and the size of the galaxy within
the framework of the merger hypothesis.

\citet{Ebisuzakietal1991} proposed that the formation of a black
hole binary might solve this difficulty. If both progenitors in a
collision between two galaxies contain central massive black holes,
these black holes would form a binary as suggested by
BBR. The back reaction of the sinking of black holes through dynamical
friction and the subsequent hardening of the black hole binary cause a heating of the
stars in the core, leading to an expansion of the
core. Using $N$-body simulations of spherical galaxies with and without
central black holes, they demonstrated that the core size, defined as
the density-weighted distance from the density center
\citep{CasertanoHut1985}, increases in the case of a merger of two galaxies with central
black holes.

In the mid-1990s, high-resolution observations by HST revealed that the
``cores'' of the giant elliptical galaxies are not really cores with a
flat volume density, but rather very shallow cusps in which the
density continues to rise toward the center as $\rho \propto
r^{-\alpha}$, where $\rho$ is the volume luminosity density and $r$ is
the distance from the center, with power index $\alpha = 0.5 \sim 1$
\citep{Gebhardtetal1996,Byunetal1996}.

Around the same time, the completion of the GRAPE-4 \citep{Makinoetal1997}
made it possible to perform $N$-body simulations of merging galaxies
with central black holes using far more particles than had been
previously possible. For example, \citet{MakinoEbisuzaki1996} performed a
simulation of repeated mergers of galaxies with central black holes,
where the final product of one merger simulation is used as the
progenitor for the next simulation. They found that the central structure of
the merger remnant converges to a unique profile, in the form of a central
cusp around the black hole with a slope of approximately $-0.5$; the
total mass of the stars in the cusp region is comparable to the mass of the
black hole binary. Furthermore, \citet{NakanoMakino1999a,NakanoMakino1999b} showed,
by a combination of $N$-body simulations and analytic arguments,
that this shallow cusp can be explained by the fact that the distribution
function of stars has a lower cutoff energy $E_0$. If there is a lower
cutoff in energy distribution, one can show in a simple derivation
that the density profile close to the central black hole must
have a slope with a power of $-0.5$.  Such a cutoff is therefore naturally formed when
black holes sink to the center and form a binary.

In this theory, the radius of the cusp region relative to the
effective radius of the galaxy is proportional to the mass of the
black hole relative to that of the parent galaxy. Thus, the
correlation between the spheroidal mass and the black hole
mass\citep{Magorrianetal1998} nicely explains the correlation between
the cusp radius and the effective radius.  Indeed,
\citet{Milosavljevicetal2002} estimated ``the mass ejected from the
central cusp'' of observed ellipticals, and found that it correlates
well with the black hole mass.  The correlation between the spheroidal
mass and the black hole mass itself is thus explained by the merger scenario.

\subsection{The Fate of a black hole binary}

As we reviewed in the previous section, a merger
of galaxies with central black holes can account for the observed
characteristics of the central structure of elliptical galaxies quite
well. However, we are still faced with one fundamental question: what will
happen to the binary black hole?

Once formed, a binary black hole further hardens through
interactions with nearby stars, much in the same way as hard
binaries in globular clusters continue to harden
\citep{Spitzer1987,HeggieHut2003}. If we can assume that the stellar
density around the black holes is roughly constant in time, the
hardening rate of the binary, $dE_b/dt$, is also constant. 

However, there are two significant differences. One is that the black
hole binary is far more massive than the field stars. Therefore,
it stays at the center of the galaxy with only a very small random
velocity. The second is that a galaxy contains far more 
stars than globular clusters do, and therefore the central two-body
relaxation time is much longer for a galaxy than for a globular
cluster. BBR argued that because
of these two differences, stars which can interact with the binary
will sooner or later be depleted, and the hardening rate of the black
hole binary will drop to practically zero. Once the nearby stars
(stars with sufficiently small angular momentum to approach to the
black hole binary) are scattered into different orbits, in other words, after the ``loss
cone'' is depleted, the hardening rate drops to a quite small value. In
this stage, the hardening rate is determined by the timescale
in which the loss cone is refilled by thermal relaxation.  Thus, in
real ellipticals with very long central relaxation times, the timescale of
refilling of the loss cone is very long and therefore the hardening timescale of the binary
would be much longer than the Hubble time.

Whether or not this loss-cone depletion actually occurs has been one
of the key issues in the theoretical and numerical studies of massive
black hole binaries. Early simulations
\citep{Makinoetal1993,MikkolaValtonen1992} could not cover a large
enough range for the number of particles (and thus for the relaxation time) to see the
corresponding change in the hardening timescale. \citet{Makino1997} performed merger
simulations with a number of particles $N$ in the range of 2K to 256K,
and found that the hardening timescale showed a clear increase, for increasing $N$.
 From numerical results, Makino estimated that the
hardening timescale was proportional to $N^{1/3}$. This dependence is
notably weaker than BBR's theoretical prediction that the hardening
timescale should be proportional to the relaxation time, or
$N$. \citet{QuinlanHernquist1997} performed similar calculations with
up to 200K stars, and found a qualitatively similar dependence of the
hardening timescale on the number of stars. However, they observed
that the evolution of the black hole binary was almost the same for the
runs with 100K stars and the run with 200K stars, and concluded that
for large enough $N$ the hardening rate would become independent of
$N$. \citet{Chatterjeeetal2003} repeated similar calculations as
performed by \citet{QuinlanHernquist1997}, but using up to 400K stars,
and reached the same conclusion. Both \citet{QuinlanHernquist1997} and
\citet{Chatterjeeetal2003} used a combination of the SCF technique
\citep{HernquistOstriker1992} and direct calculation, where
interactions between field stars are treated using SCF, while
\cite{Makino1997} relied on a fully direct calculation of all forces using the
GRAPE-4. Though the SCF approximation does not eliminate two-body relaxation
\citep{HernquistBarnes1990,HernquistOstriker1992}, it may have
affected the dependence of the relaxation timescale on $N$ in a
complex way.

\citet{MilosavljevicMerritt2001} performed simulations of mergers with
central black holes, by again using a composite method, but here
composite in the time domain and not in the spatial domain. Before the
formation of the black hole binary, they employed a parallel treecode with individual
timesteps \citep{Springeletal2000}. Just before the binary formation,
they then switched over to a direct calculation using a general-purpose parallel
computer. The use of a general-purpose parallel computer for direct
calculations limited the number of particles to 32K
and less, but their simulation is unique in that they started from
a self-consistent model with a central cusp of slope $-2$. They did not
find any dependence of the hardening rate on the number of stars.

To summarize,  there is no single accepted view
for the evolution of massive black hole binaries in the center of
galaxies. The simple analytic theory (BBR) predicts that the hardening
timescale should be determined by the timescale for refilling the loss-cone,
and therefore should be proportional to the relaxation
time, or roughly speaking, the number of stars in the system. The
results of numerical studies range from no dependence
\citep{MilosavljevicMerritt2001} to a dependence
$\propto N^{1/3}$\citep{Makino1997}.

In this paper, we  give a clear and decisive answer to the
question  whether the hardening timescale depends on the number of
particles used in the simulation, and if so, in what way. In section 2 we
describe the initial model and the numerical method used. In section 3
we give the result.  Section 4 contains the discussion.

\section{Numerical Method and  Initial Models}

\subsection{Numerical method}

We performed $N$-body simulations of galaxies with massive black
holes. For all calculations, we used a program with direct force
calculation and fourth-order Hermite integrator with individual
(block) timestep \citep{MakinoAarseth1992}. For gravitational
interaction between field stars, we apply the usual Plummer
softening. The size of the softening is described in section
\ref{sect:initialmodels}. The gravitational interactions between black
holes and that between black holes and field particles are calculated
with very small softening, $\epsilon_{BH} = 10^{-6}$. Since we do not
apply any regularization technique, we need to guarantee that the
gravitational force does not diverge.  

For calculation of gravitational forces from field particles (to both
the field particles and black holes), we used the special-purpose
hardware GRAPE-6\citep{Makinoetal2002}. The calculation of the forces
from black holes was done on the host computer to maintain sufficient
accuracy, in a way similar to \cite{MakinoEbisuzaki1996}. The
relative accuracy of the pairwise force calculated with ``High
accuracy'' versions of GRAPE hardwares (GRAPE-2, 4 and 6) is roughly
single precision (24-bit mantissa). This accuracy is usually
sufficient, since the errors of the forces from many particles
partially cancel each other. However, single-precision is not quite
enough for forces from black hole particles. The reason is that some
particles, and most notably the black hole particles themselves after they
formed a binary, orbit around a black hole particle a number of times,
essentially feeling only the force from one black hole particle. In this case, the
round-off error accumulates, and the error in the total energy becomes
alarmingly large. Moreover, all the errors are generated from the black hole
binary and stars which are close to them, the behavior of which we are
interested in. In order to guarantee sufficient accuracy, we
chose to calculate all gravitational forces from black hole particles on the
host computer, using full double-precision (53-bit mantissa) numbers.

No relativistic effect was taken into account. Our calculation is
purely Newtonian. We do not model the accretion to black holes or collisions
between stars, since their cross sections are small.

\subsection{Initial Models}
\label{sect:initialmodels}

In this paper, we consider a simple model, where we place two massive
point-mass particles in a spherical galaxy. For our standard set of
runs, the initial galaxy model is a King model with nondimensional
central potential  $W_0=7$. We use Heggie units, where the mass
$M$ 
and the virial radius $R_v$ of the initial galaxy model and gravitational
constant $G$  are all unity. In these units, the binding energy of the
initial galaxy is $E = -1/4$.

The mass of the black hole particles is
$M_{BH}=0.01$. They are initially placed at $(\pm 0.5,0,0)$ with velocity
$(0,\pm 0.1,0)$. Thus, they are initially outside the core, at the
apocenter of nearly radial orbits.

We varied the number of particles $N$ from 2,000 to 1,000,000. To
investigate the possible dependence of the results on the softening
length, we tried three different choices for the softening length: (a)
$\epsilon=0.01$, (b) $\epsilon=0.01/(N/2000)^{1/3}$, (c)
$\epsilon=20/N$. All give the same $\epsilon=0.01$ for $N=2000$, but
they have a different dependence on $N$.

The largest calculation (1 million particles for up to $t=300$) took
about one month on a single-host, 4-processor-board GRAPE-6 system
with a peak speed of 4 Tflops. The total number of individual timesteps
was $1.2\times 10^{11}$. In other words, the (harmonic) average timestep
is around $2.5\times 10^{-3}$. In comparison, the total number of block
timesteps is $2.1\times 10^8$. Thus, the typical timestep size
of the particle with the smallest timestep (generally the black hole
particles themselves or particles close to them) is $1.4 \times
10^{-6}$, more than 1000 times smaller than the average stepsize.
Without the use of individual timesteps, this calculation would have
required a fraction of a Petaflops-year, rather than 1/3 Teraflops-year.

For all calculations, the total energy is conserved to better than
0.1\%, or better than 1\% of the binding energy of the BH binary. We
did several test calculations with both higher and lower accuracy
criteria, but found no systematic difference in the final results.

\section{Results}

\subsection{The $N$-dependence}

\begin{figure}
\plotone{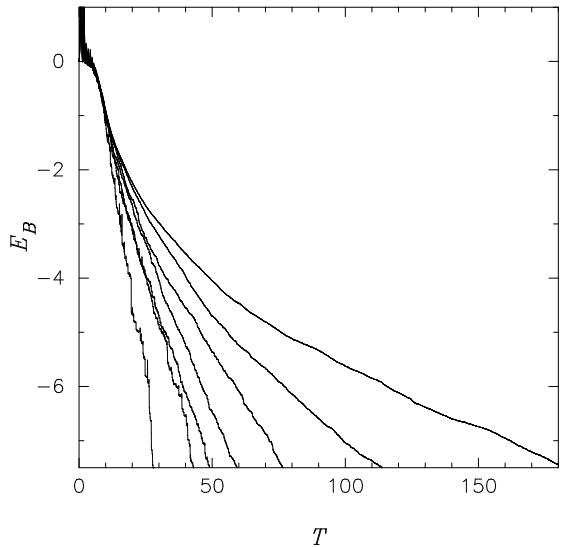}
\caption{The evolution of the specific binding energy $E_b$ of the
binary black hole. Curves give the results for $N=10K$, 20K, 50K, 100K,
200K, 500K and 1M (left to right). For all calculations, the softening
length is $\epsilon=0.01$.
\label{fig:ebfign}}
\end{figure}

Figure \ref{fig:ebfign} shows the time evolution of the specific
binding energy (per unit of reduced mass) of the black hole binary. The
relation between the semi-major axis $a$ and the binding energy is
\begin{equation}
a = - \frac{M_{BH}}{E_b}= - \frac{1}{100E_b}.
\end{equation}

The relation between the orbital velocity of the black holes and $E_b$ is
given simply by
\begin{equation}
v_{BH}= v_c\sqrt{|E_b}|,
\end{equation}
where
$v_c$ is the three-dimensional velocity dispersion of the field
stars, in a central region of the galaxy, chosen to be large enough not to be
affected significantly by the black holes.  Thus, if we interpret the host galaxy as a galaxy with a
velocity dispersion of 300 km/s, the black hole binary with $E_b=-1$
has an orbital velocity of 300 km/s.

 From figure \ref{fig:ebfign}, it is clear that the evolution timescale
continues to depend on $N$ for the entire range of $N$ for which we performed
our simulations, with no indication whatsoever of even an onset toward convergence.

\begin{figure}
\plotone{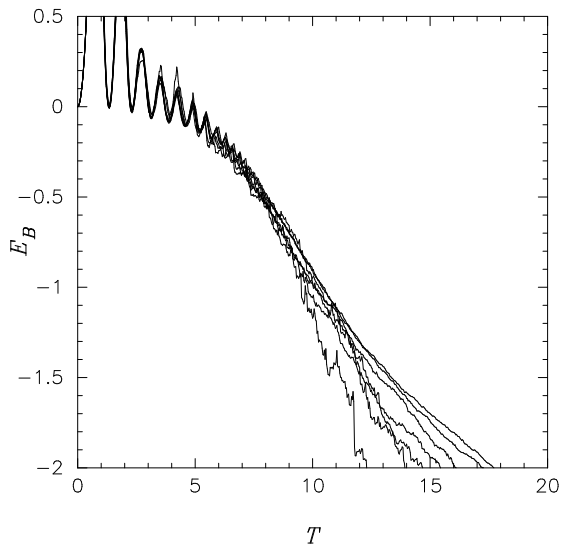}
\caption{Same as figure \protect \ref{fig:ebfign}, but only the early
evolution is shown.
\label{fig:ebfign2}}
\end{figure}

Figure \ref{fig:ebfign2} shows the early evolution. Initially the
binding energy shows large oscillations, simply because the black hole
particles are not bound to each other but orbit within the parent
galaxy in highly eccentric orbits. As their orbits shrink through
dynamical friction, the amplitude of the oscillations becomes smaller,
and eventually the two black holes become bound. We can see that the early evolution
of the black hole binary, before the specific binding energy reaches
$-1$, is almost independent of $N$. However, after $E_b$ reaches $-2$,
the evolution timescale shows a strong dependence on the number of
particles.

\begin{figure}
\plotone{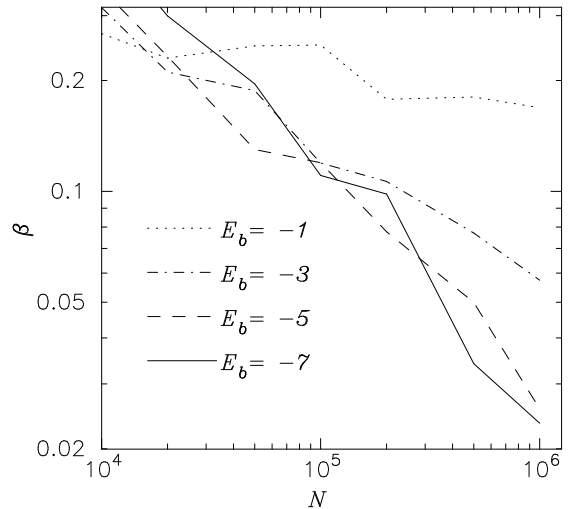}
\caption{The hardening rate $\beta = |\Delta E_b/\Delta t|$ plotted as a
function of the number of particles $N$. Dotted, dash-dotted, dashed
and solid curves denote the values measured from the time at which
$|E_b|$ reached 1, 3, 5 and 7, respectively, until the time
at which $|E_b|$ has increased by an additional amount of $+0.5$.
\label{fig:defig}}
\end{figure}

To quantitatively evaluate the dependence of the hardening rate on the
number of particles, we calculate the hardening rate $\beta$, defined
as
\begin{equation}
\beta = |\Delta E_b/\Delta t|.
\end{equation}
Here, $\Delta t = t_1 -t_0$ where $t_0$ and $t_1$ are the times at which
$E_b$ reached the values $E_{b,0}$ and  $E_{b,0}+\Delta E_b$,
respectively. We use $\Delta E_b=-0.5$ for all
values of $E_{b,0}$.  Figure \ref{fig:defig} shows the
result, for $E_{b,0}=-1,-3,-5,-7$. When $E_b=-1$, the hardening rate is almost
independent of $N$. However, as the binary becomes harder, $\beta$
decreases, and the decrease is larger for larger $N$. Thus, the
hardening rate $\beta$ for large values of $|E_b|$ shows a strong
dependence on the number of particles $N$.

This result is exactly what is expected from the simple loss-cone
argument: after the loss cone is
depleted, $\beta$ should be inversely proportional to the relaxation
time, which is proportional to $N$. Note that we used a constant
softening, for which the Coulomb logarithm does not depend on $N$.

\begin{figure}
\plotone{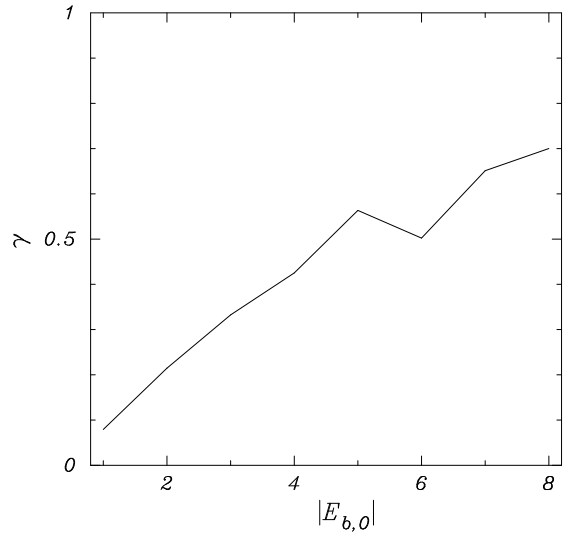}
\caption{The slope $\gamma$ of the dependence of the hardening rate
$\beta$ on the number of particles $N$, plotted as a function of the
binding energy $E_{b.0}$ at which $\beta$ is measured.
\label{fig:deslope}}
\end{figure}

If the loss-cone argument would be correct, writing
$\beta \propto N^{-\gamma}$ would let $\gamma$ approach unity for
large enough $N$.
Figure \ref{fig:deslope} shows $\gamma$, calculated
using the value of $\beta$ for $N=20{\rm K}$ and 1M. The choice of
these values of $N$ is somewhat arbitrary, but as one can see from
figure \ref{fig:defig}, we obtained similar figures with different
choices of $N$. We could use a least-square fit, but decided against
it since there is no obvious reason to assign equal weights to results
with different $N$. 

We can see that
$\gamma$ indeed increases as $|E_b|$ increases. Even at
$E_b=-8$, $\gamma$ has not converged to a final value. If we could extend
the calculation to higher values of $E_b$, we would be able to
determine whether or not $\gamma$ really approaches unity.
Unfortunately, it would be too time consuming, even on a GRAPE-6, to
further extend the calculations, given that the hardening rate is so small for
large $N$.

 From the current simulations, we can safely conclude that the
hardening rate $\beta$ depends on the number of particles $N$, and the
power index of the dependence $\gamma$ numerically obtained is larger
than 0.7. The numerical result is consistent with the simple loss-cone
argument, which predicts $\gamma=1$.

\subsection{Dependence on the softening length}

\begin{figure}
\plotone{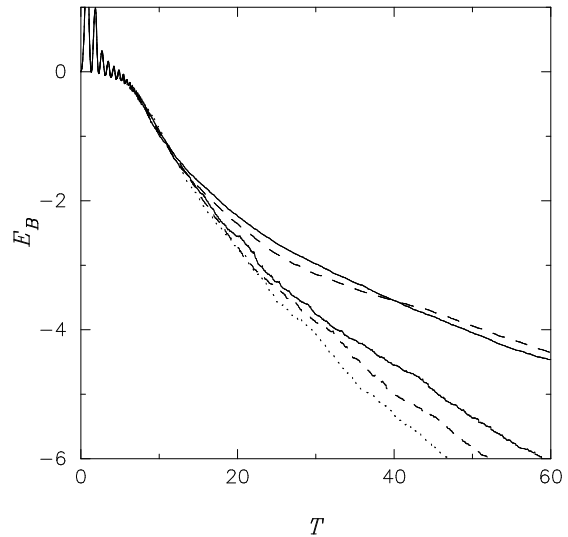}
\caption{The evolution of specific binding energy $E_b$. The curves denote,
from top to bottom, $N=10^6, \epsilon=0.01$ (solid), $ \epsilon=0.00045$
(dashed), $N=2\times 10^5, \epsilon=0.01$ (solid), $ \epsilon=0.0001$
(dashed), and $ \epsilon=0.00001$ (dotted).
\label{fig:ebeps}}
\end{figure}

The results given in the previous subsection demonstrate clearly that
the hardening rate depends on the number of particles. In this and
following subsections, we will check whether this result is really reliable.
First we look at the effect of the softening. The relatively
large and constant softening used in our standard runs has the effect
of suppressing  two-body relaxation. In particular, in the core of
the galaxy, the effective Coulomb logarithm might be very small,
resulting in unphysical suppression of the relaxation effect. Since
the timescale of the loss-cone refill is related to the relaxation
timescale, it is crucial to express the relaxation with a reasonable
accuracy. To test the effect of the softening, we performed several
runs with a much smaller softening length. Figure \ref{fig:ebeps} shows
the result. For $N=2\times 10^5$, reducing $\epsilon$ by a factor of
100 resulted in only a small increase in the hardening rate. Furthermore, in the
case of runs with $N=10^6$, the change in the softening has
practically no effect on the hardening rate. Thus, we can safely conclude
that the effect of the softening is, if any, sufficiently small that
it does not affect the results in the previous section.

\subsection{Dependence on the initial models}

Before the loss cone is depleted, the hardening rate is expected to be
proportional to the central density of the parent galaxy. After the
loss cone is depleted, the hardening rate is determined by the timescale
at which the stars close to the loss cone diffuse into the cone. Thus,
here again, the hardening timescale is expected to depend on the initial
central density.

\begin{figure}
\plotone{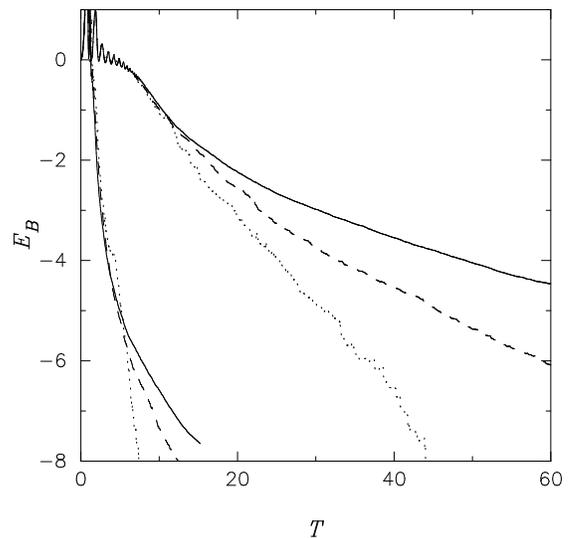}
\caption{The evolution of specific binding energy $E_b$ for runs with
different initial galaxy models. Solid, dashed and dotted curves display the results
for $N=10^6$, $2 \times 10^5$, and $2\times 10^4$. The left three curves
are for a King model with $\psi_0=11$, while the right three curves are for
$\psi_0=7$. 
\label{fig:ebphi}}
\end{figure}

Figure \ref{fig:ebphi} shows the results of runs with different
initial galaxy models. As expected, the hardening is faster for models
with deeper central potential (higher central density). However, for
all runs the hardening rate depends  on the number of particles.

\begin{figure}
\plotone{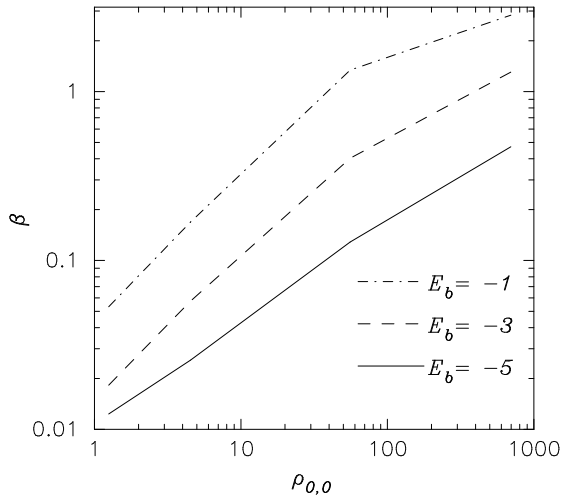}
\caption{The growth rate $\beta = |dE_b/dt|$ plotted as a function of
the initial central density $\rho_{0,0}$ of the parent
galaxy. Different curves denote the growth rate at different values of
$|E_b|$. 
\label{fig:phide}}
\end{figure}

Figure \ref{fig:phide} shows the growth rate $\beta$ as a function of
the initial central density of the parent galaxy, $\rho_{0,0}$. In 
the early phase ($E_b=-1$),  $\beta$ is roughly proportional
to $\rho$ for $\rho < 10^2$. For higher $\rho$ or for later phases the
dependence is somewhat weaker, presumably because the black hole
binary already has ejected nearby stars, thereby reducing the central density.

\subsection{Loss cone depletion and wandering of the binary}

In the previous subsections we have seen that the behavior of the
hardening rate is consistent with the loss cone argument. In this
subsection, we directly investigate whether or not the loss cone is actually
depleted.

\begin{figure}
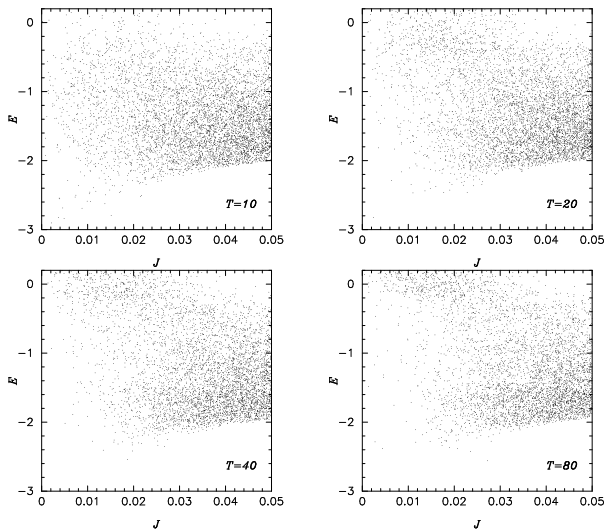

{\plottwo{f8a.eps}{f8b.eps}}

{\plottwo{f8c.eps}{f8d.eps}}

\caption{The distribution of particles in the $(E,J)$ plane at times $T=10$, 20, 40 and 80 (from top left to
bottom right). The number of
particles is $10^6$.
\label{fig:ej1M}}
\end{figure}

Figure \ref{fig:ej1M} shows the distribution of particles in the $(E,J)$
plane, where $E$ is the specific energy and $J$ is the specific total
angular momentum.  We use the coordinate origin as the reference point
for the angular momentum. We also tried to use the center of mass of
the black hole binary, but that resulted in practically indistinguishable
figures.  Here, we can clearly see that the number of particles with $J<0.01$
is more and more depleted as time proceeds. At $T=10$, only particles with low energy
($E<-2$) are affected. However, the depletion reaches higher energy levels as
time proceeds, and by $T=80$ only stars with nearly zero energy 
are left with low angular momentum. Note that there were no particles with nearly zero
energy at $T=10$. These barely bound particles have been kicked to
high-energy, long-period orbits by the binary black hole.

\begin{figure}
\plotone{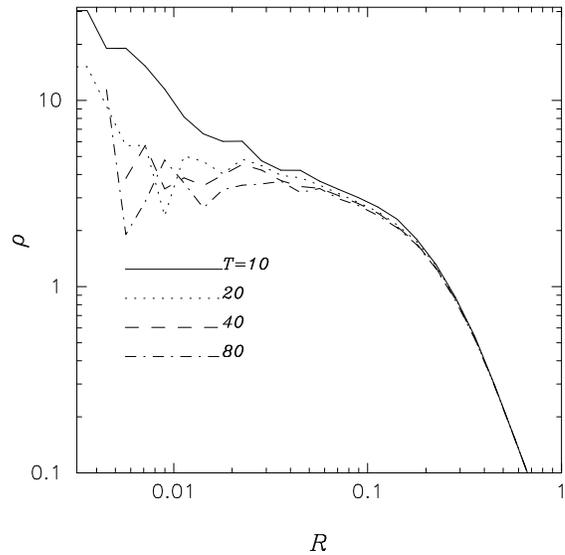}
\caption{Radial density profiles at times 10, 20, 40 and 80 for
the run with $N=10^6$. 
\label{fig:prof1M}}
\end{figure}

Figure \ref{fig:prof1M} shows the radial density profiles for the same
snapshots as used for Figure \ref{fig:ej1M}. Here we took the center
of mass of the black hole binary as the center of the coordinate.
Though the loss cone depletion is clearly visible in the distribution
of particles in the $(E,J)$ plane, the density profiles  show no
clear sign of the existence of a loss cone. This result is not at all surprising. If
the distribution of particles in the $(E,J)$ plane is the same as that of
the initial King model, adding the central black hole potential
would result in a density cusp with a slope of $-0.5$
\citep{NakanoMakino1999b}. Thus, the fact that the density does not
increase toward the center actually means that the particles close to
the black hole binary are strongly depleted.

\begin{figure}
{\plottwo{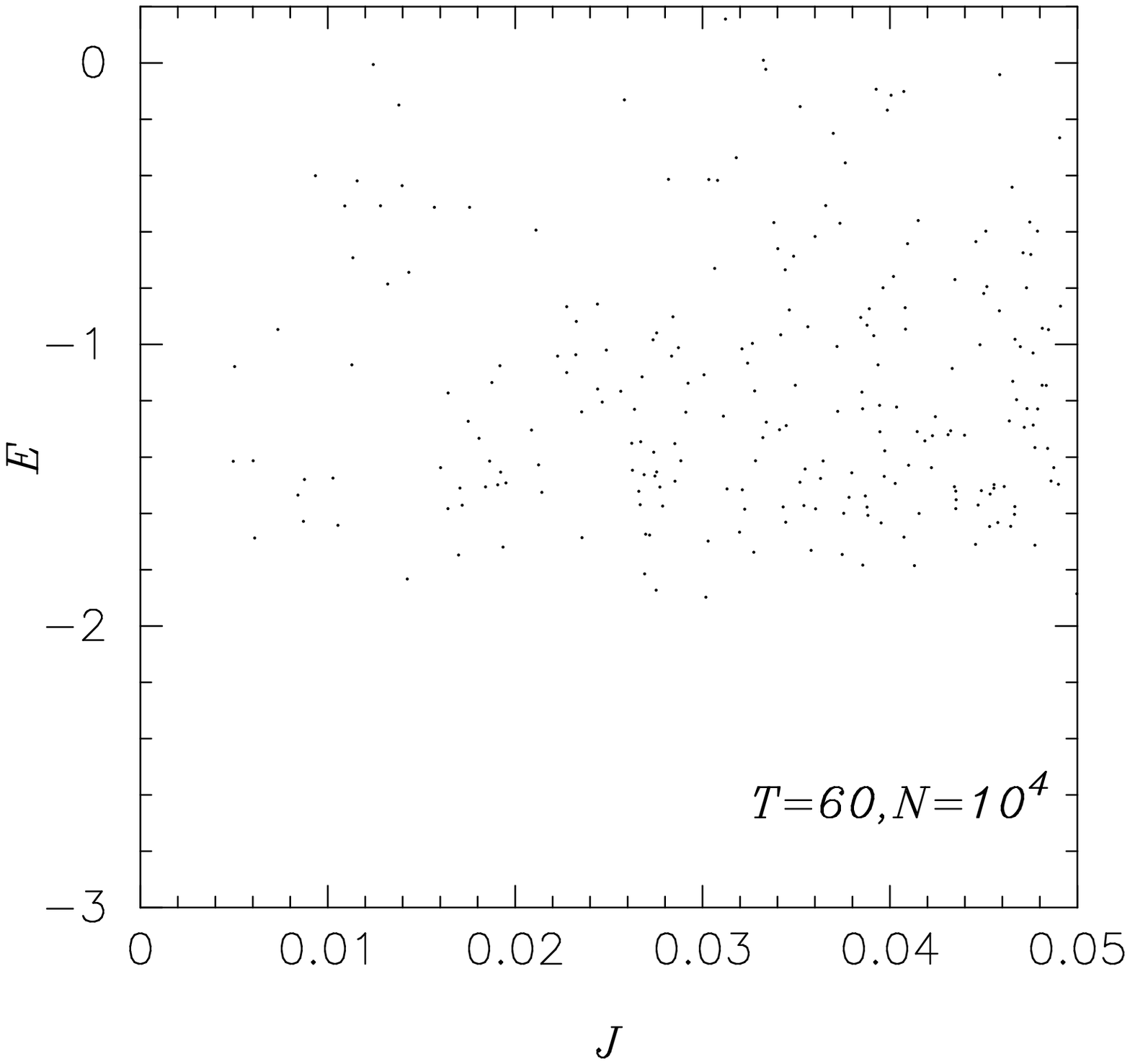}{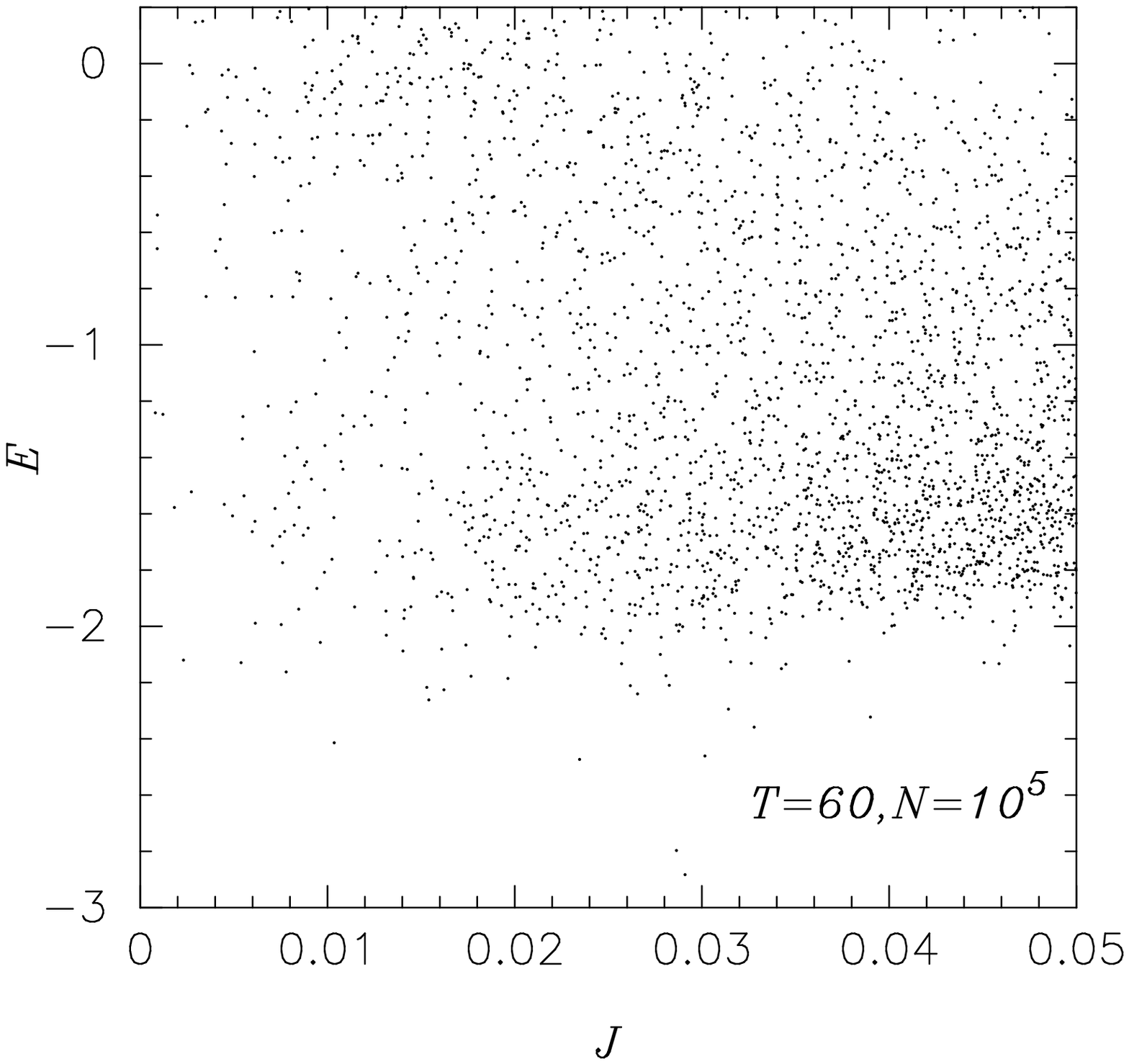}}

\caption{The distribution of particles in the $(E,J)$ plane for runs with
$10^4$ (left) and $10^5$ (right) particles, at $t=60$.
\label{fig:ejn}}
\end{figure}

Figure \ref{fig:ejn} shows the distribution of particles in the
$(E,J)$ plane for runs with smaller number of particles. The
distribution for  $N=10^4$ does not show any sign of
depletion, while the distribution for $N=10^5$ shows a weak indication
of depletion. It also show an
enhancement of nearly zero-energy stars with small angular
momentum.

\begin{figure}
\plotone{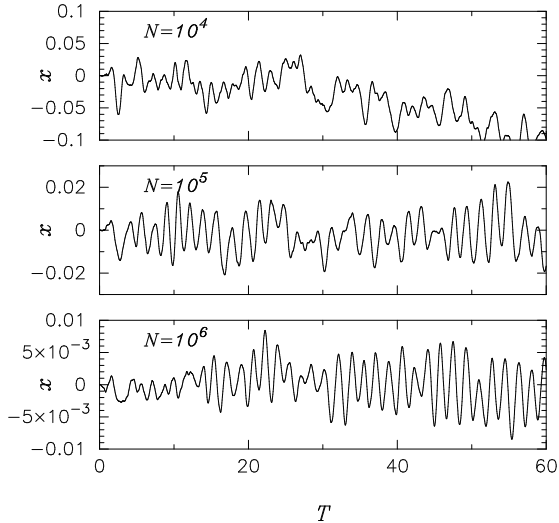}
\caption{The time variation of the $x$ component of the center of mass
of the black hole binary. The number of particles is $10^4$(top),
$10^5$(middle), and   $10^6$(bottom). Note that the vertical axis is
scaled in proportion to $1/\sqrt{N}$. 
\label{fig:cmtrak}}
\end{figure}

Note that figure \ref{fig:ej1M} shows that the periastron
distance of the particles that are depleted, a measure for the effective radius
of the loss cone, is 0.01 or larger. This radius is  much bigger than
the semi-major axis of the binary, which is around 0.001 at $T=80$.
The reason why the effective size of the loss cone is much bigger than
the semi-major axis of the black hole binary is the wandering of the binary. Figure
\ref{fig:cmtrak} shows
the time variation of the $x$ coordinate of the center of mass of the
binary. The typical wandering distance is roughly proportional to
$N^{-1/2}$, as theoretically predicted and demonstrated  by previous
numerical work \citep{Makino1997,MilosavljevicMerritt2001}. For
$N=10^6$, the typical wandering distance is around 0.01, roughly
consistent with the size of the loss cone.

\section{Discussion}

\subsection{Comparison with previous works}

As we summarized in the introduction, several researchers have performed
$N$-body simulations of the evolution of massive black hole
binaries in the center of a galaxy, and nobody has obtained a result
which could be interpreted as being consistent with the simple
loss-cone argument by \citet{Begelmanetal1980}. In the present work,
however, we have obtained a
result that is consistent with the loss cone argument. In order to
understand the cause of the discrepancy, we first discuss our own previous work
\citep{Makino1997}, and then work by others
\citep{QuinlanHernquist1997,MilosavljevicMerritt2001,Chatterjeeetal2003}.

\citet{Makino1997} obtained a value for the slope $\gamma$ of the
dependence of the binary hardening rate $\beta$ on the number of
particles $N$ of $\gamma \sim 1/3$, from merger simulations
with $N=2,048$ to 262,144. If we compare his figure 1 and our figures
\ref{fig:ebfign} and \ref{fig:ebfign2}, the reason why the value of
$\gamma$ was small is obvious: the value of the binding energy at which
$\beta$ was measured was simply too small. Therefore $\beta$ had not yet
reached the value determined by the relaxation timescale. As shown in
figures \ref{fig:defig} and \ref{fig:deslope}, $\gamma$ increases as
$|E_b|$ increases. In hindsight, it is clear that previous simulations
did not cover a long enough time.

\citet{QuinlanHernquist1997} performed simulations very similar to
those presented here. They found that $\beta$
depended on $N$, for $N<10^5$. However, the hardening rate was
practically the same for the runs with $N=10^5$ and $N=2\times
10^5$. They explained this result as follows. In their calculation,
they had used the SCF method to evaluate the gravitational interaction
between field stars. Therefore the two-body relaxation had been
strongly suppressed, and the dependence of the hardening rate on the
number of particles should come only through the wandering (Brownian
motion) of the black hole binary.
The random velocity of the black hole binary would be proportional to
$\sqrt{N}$. However, the distance the wandering covers would not
become arbitrary small, since the black hole binary would deplete the
``loss cone'' and create a kind of vacuum around it, the black hole binary would
not feel a restoring force as long as it remained in the vacuum.

The above explanation looks plausible, but unfortunately 
\citet{QuinlanHernquist1997} gave no evidence that the distance
covered by the black hole binary became independent of $N$. Our
simulations demonstrated that the excursion distance is proportional
to $1/\sqrt{N}$, for up to $10^6$ particles.

An additional complication with their calculations is that it is difficult to
estimate the effects of two-body relaxation and wandering,
because they used a mass for the particles which depended on their
initial angular momenta (or, precisely, the periastron
distances). They used this trick to increase the mass resolution near
the center of the galaxy. However, because of this trick, the typical mass
of stars that interact with the black hole binary would become larger
as the black hole binary kicks out more and more of the nearby (low-mass)
particles. This implies that the strength of the Brownian motion depends on
how much mass has been kicked out of the core. Also, even though the effect
of direct two-body encounters of field particles is suppressed, these particles can still
indirectly exchange energy and angular momentum since all field stars
directly interact with the black hole binary which has a finite random
velocity.  Even distant massive particles do interact significantly with the black
holes. 
Thus, the way the two-body relaxation scales is quite
difficult to evaluate for their scheme.

Quinlan and Hernquist tried one simulation in which the center of mass of the binary is
fixed at the origin of the coordinates, and found that the hardening rate
dropped dramatically. They argued that this drop is evidence for
wandering being important. However, there is another, equally
plausible explanation. In their SCF expansion, they incorporated only
spherically symmetric terms. Thus, when they fixed the center of
mass of the binary to the origin, the gravitational potential
calculated became strictly spherically symmetric, except for the
multipole moment of the binary black hole. Thus, the angular momentum
of field stars with periastron distance larger than the semi-major
axis of the binary is conserved. In other words, the loss cone, once
depleted, can never be refilled. Once they allow the center-of-mass
motion of the binary to occur, however, the gravitational potential is no
longer spherically symmetric, and angular momenta (as well as
energies) of field particles change through interactions with the
center-of-mass motions of the binary. This process effectively works
as a relaxation mechanism and causes the refilling of the loss cone.

To summarize, though \citet{QuinlanHernquist1997} observed that in
their calculation the hardening rate became independent of $N$ for
large $N$, it is somewhat difficult to generalize their result,
obtained with a spherically-symmetric potential expansion code and
radius-dependent particle mass, to real $N$-body system or real
elliptical galaxies.

\citet{Chatterjeeetal2003} performed calculations similar to that
presented in \citet{QuinlanHernquist1997}, using essentially the same
calculation code. However, they did use a few non-spherical terms in
their potential expansion,
and they used equal-mass particles instead of particles having radius-dependent
mass. Thus, their results can be directly compared with our results
without much complication. In the text, they stated that the
hardening rate became constant for the value of $N$ around $2\sim 4
\times 10^5$.
For their 400K run,
$M_{BH}=0.00125$, which is 1/8 of the value we used.
Unfortunately, the simulation was stopped before we would expect to 
to see whether the loss-cone depletion would or would not occur.


Unlike the previous two papers, \citet{MilosavljevicMerritt2001}
calculated the actual merger of two galaxies with central black holes,
as \citet{Makino1997} did. The difference between 
\citet{MilosavljevicMerritt2001} and \citet{Makino1997} is that the
former started with a galaxy model with a density cusp with $\rho
\propto r^{-2}$ around the central black hole, while the latter used a
King model with a finite core as the initial model. For simulations of
mergers of elliptical galaxies with massive central black holes ($10^8
\msun$ or larger), a galaxy model with finite-size core would be more
appropriate, since the shallow ``cusp'' of such large ellipticals
corresponds to a cutoff of the distribution function at finite
energy. The stellar distribution in the central region of spiral galaxies
is consistent with a cusp of $\rho
\propto r^{-2}$. Therefore, for simulations of mergers of spiral galaxies,
the initial model used by \citet{MilosavljevicMerritt2001} may be more
appropriate.

\citet{MilosavljevicMerritt2001} performed three runs with 8K, 16K and
32K stars, and found that the hardening rate was independent of the
number of particles.  As discussed in their paper, this result was
stemmed from the fact that the loss cone was not depleted in their simulations,
because of two reasons. The first is that the initial central density
of their model was high, since they tailored the distribution function
so that the progenitor galaxies had central cusps around the black
holes. This is a situation very different from that used in
other studies, where black holes were placed off-center in a single
galaxy or placed at the centers of galaxies with relatively large
cores. Since the initial stellar density around the black holes is
very high, the binary initially hardens very rapidly. A binary with a
small semi-major axis has a small interaction cross section, which implies
that it takes a long time to deplete the nearby stars. The second
reason is the relatively small number of particles employed in their
simulation, which resulted in rather large random velocities. Therefore,
Brownian motion of the binary covered a fairly large radius. Even in
their largest calculation with 32K particles, the total mass of the
stars which can enter the region covered by the Brownian motion of
the black hole binary is much larger than the mass ejected by the
binary. Thus, the very high initial central density and the relatively
small number of particles conspired together to prevent the loss cone from being
depleted.

If \citet{MilosavljevicMerritt2001} could have used a much larger number of particles, the wandering
distance would have shrunk, and the loss cone would have been formed
early on. According to their own estimate
\citep{MilosavljevicMerritt2001},  for $N>2\times 10^5$ the loss cone
depletion becomes important.

To summarize, there is no real discrepancy among the results of full
$N$-body simulations. \citet{Makino1997} observed weaker dependence of
the hardening rate than obtained in the present study, simply because
his calculations were not long
enough. \citet{MilosavljevicMerritt2001}  found no dependence on $N$,
essentially because the number of particles they used was too small
for the loss cone to be depleted.  

\subsection{Merging timescale of massive black hole binaries in
ellipticals}

As first suggested by BBR and confirmed by a number of followup works,
if the hardening timescale of the black hole binary is proportional to
the relaxation timescale of the parent galaxy, the evolution timescale
of  typical  black hole binaries in  elliptical galaxies exceeds the
Hubble time by many orders of magnitude. In other words,  binaries are
unlikely to merge through  encounters with field
stars and gravitational wave radiation.

Our results strongly suggest that the hardening timescale is indeed
determined by the relaxation timescale, for large enough $N$ and after
the binary becomes sufficiently hard. This implies that
gravitational interaction with field stars is insufficient to let the
binary merge.

There are a number of alternative mechanisms that may lead to the merger
of the black hole binary. If there is a significant amount of gas left
at the center, or if gas is supplied from the disk during the merger
event, it would certainly change the entire picture. However, in the case of a
merger of two ellipticals, there is not much cold gas left in the
resulting galaxy. In this case, the most likely outcome is that the binary,
stuck at a certain semi-major axis, stays at the center of the loss
cone.

If the eccentricity of the binary goes up, the timescale of orbital
evolution by gravitational wave radiation is reduced
significantly. Roughly speaking, if the eccentricity reaches 0.9, a fair
fraction of the binary black holes would merge in a time less than
the Hubble time. Some of the early $N$-body simulations and scattering
experiments have focused on this possibility \citep{Makinoetal1993,MikkolaValtonen1992}. However, the general consensus
seems to be that the eccentricity does not change much during the
hardening. 

This situation can change if there are more than two massive black
holes. If the binary black hole has a long lifetime, it is quite natural
to assume that some of the  galaxies which contain binary black
holes will undergo a further merger with another galaxy with a central massive
black hole or a binary. If we regard the black holes as
point-mass particles interacting through Newtonian gravity, then with three
(or more) black holes we expect at most one black hole binary to be
left in the galaxy, having ejected all other black holes by the gravitational slingshot
mechanism \citep{SaslawValtonenAarseth1974}. However, here the 
eccentricity effects might play an important role. Simple estimates assuming a
thermal distribution of eccentricities \citep{MakinoEbisuzaki1994}
suggest that, during repeated three body interactions, the
eccentricity of the binary can reach a very high value, resulting in 
quick merging through gravitational wave radiation. In principle, it
is possible that multiple black holes form a stable hierarchical
system, where evolution of the outer binary is halted because of the
loss cone depletion. In this case, however, the inner binary would
typically have a semi-major axis of orde 1/10 of that of the outer
binary, and the gravitational wave radiation timescale of the inner
binary would generally be short. Also, the Kozai mechanism could play a role
in increasing the eccentricity of the inner binary
\citep{Blaesetal2002}.

So far, our primary attention has been directed to supermassive black holes in large
ellipticals, since observational evidence for massive dark objects is
strongest for large ellipticals. If LSB galaxies had central black holes
(which would be more like intermediate-mass black holes than massive
black holes), some of them must have experienced major merger events and
therefore are likely to have contained multiple black holes at some point.
Here, again, the crucial question is whether or
not these black holes can merge. Since the black hole masses are smaller, if the
central region of LSB galaxies is dominated by normal stars, loss-cone
depletion is less effective, and black holes can become very
tight. Indeed, if we regard the field particles in our simulations as
normal stars with typical masses around a solar mass, we have performed
simulations of BH binaries with masses up to $10^4 M_{\odot}$, for
which we have seen that the
hardening rate is still pretty high. However, if the central region is
dominated by CDM, as suggested by both theory and observations, the
thermal relaxation time would be practically infinite and loss-cone
depletion would occur instantly. In the case of LSB galaxies, it is
unlikely that two black holes merge during triple interactions, because
the central potential is shallow. All black holes would be ejected
from the parent galaxy before they would get a chance to merge. Thus, unlike large ellipticals, LSB
galaxies are unlikely to display something like a Magorrian relation or
$M_{bh}$--$\sigma$ relation. 

\section*{Acknowledgements}

The authors thank Sverre Aarseth, Rainer Spurzem for discussions
related to this work, and Piet Hut for carefully reading the
manuscript.  This work is supported in part by Grant-in-aid
in Scientific Research in Priority Areas (15037203) and B (13440058)
of the Ministry of Education, Culture, Culture, Science and
Technology, Japan.

\def\mn{MNRAS}


\begin{thebibliography}{33}
\expandafter\ifx\csname natexlab\endcsname\relax\def\natexlab#1{#1}\fi

\bibitem[{{Begelman} {et~al.}(1980){Begelman}, {Blandford}, \&
  {Rees}}]{Begelmanetal1980}
{Begelman}, M.~C., {Blandford}, R.~D., \& {Rees}, M.~J. 1980, \nat, 287, 307

\bibitem[{{Blaes} {et~al.}(2002){Blaes}, {Lee}, \& {Socrates}}]{Blaesetal2002}
{Blaes}, O., {Lee}, M.~H., \& {Socrates}, A. 2002, \apj, 578, 775

\bibitem[{{Byun} {et~al.}(1996){Byun}, {Grillmair}, {Faber}, {Ajhar},
  {Dressler}, {Kormendy}, {Lauer}, {Richstone}, \& {Tremaine}}]{Byunetal1996}
{Byun}, Y.-I., {Grillmair}, C.~J., {Faber}, S.~M., {Ajhar}, E.~A., {Dressler},
  A., {Kormendy}, J., {Lauer}, T.~R., {Richstone}, D., \& {Tremaine}, S. 1996,
  \aj, 111, 1889

\bibitem[{{Casertano} \& {Hut}(1985)}]{CasertanoHut1985}
{Casertano}, S., \& {Hut}, P. 1985, \apj, 298, 80

\bibitem[{Chatterjee {et~al.}(2003)Chatterjee, Hernquist, \&
  Loeb}]{Chatterjeeetal2003}
{Chatterjee}, P., {Hernquist}, L., \& {Loeb}, A. 2003 \apj, 592, 32

\bibitem[{{Ebisuzaki} {et~al.}(1991){Ebisuzaki}, {Makino}, \&
  {Okumura}}]{Ebisuzakietal1991}
{Ebisuzaki}, T., {Makino}, J., \& {Okumura}, S.~K. 1991, \nat, 354, 212

\bibitem[{{Farouki} {et~al.}(1983){Farouki}, {Shapiro}, \&
  {Duncan}}]{Faroukietal1983}
{Farouki}, R.~T., {Shapiro}, S.~L., \& {Duncan}, M.~J. 1983, \apj, 265, 597

\bibitem[{{Ferrarese} \& {Merritt}(2000)}]{FerrareseMerritt2000}
{Ferrarese}, L., \& {Merritt}, D. 2000, \apjl, 539, L9

\bibitem[{{Gebhardt} {et~al.}(2000){Gebhardt}, {Bender}, {Bower}, {Dressler},
  {Faber}, {Filippenko}, {Green}, {Grillmair}, {Ho}, {Kormendy}, {Lauer},
  {Magorrian}, {Pinkney}, {Richstone}, \& {Tremaine}}]{Gebhardtetal2000}
{Gebhardt}, K., {Bender}, R., {Bower}, G., {Dressler}, A., {Faber}, S.~M.,
  {Filippenko}, A.~V., {Green}, R., {Grillmair}, C., {Ho}, L.~C., {Kormendy},
  J., {Lauer}, T.~R., {Magorrian}, J., {Pinkney}, J., {Richstone}, D., \&
  {Tremaine}, S. 2000, \apjl, 539, L13

\bibitem[{{Gebhardt} {et~al.}(1996){Gebhardt}, {Richstone}, {Ajhar}, {Lauer},
  {Byun}, {Kormendy}, {Dressler}, {Faber}, {Grillmair}, \&
  {Tremaine}}]{Gebhardtetal1996}
{Gebhardt}, K., {Richstone}, D., {Ajhar}, E.~A., {Lauer}, T.~R., {Byun}, Y.-I.,
  {Kormendy}, J., {Dressler}, A., {Faber}, S.~M., {Grillmair}, C., \&
  {Tremaine}, S. 1996, \aj, 112, 105+

\bibitem[{Heggie \& Hut(2003)}]{HeggieHut2003}
Heggie, D.~C., \& Hut, P. 2003, The Gravitational Million-Body Problem: A
  Multidisciplinary Approach to Star Cluster Dynamics (Cambridge: Cambridge
  University Press)

\bibitem[{{Hernquist} \& {Barnes}(1990)}]{HernquistBarnes1990}
{Hernquist}, L., \& {Barnes}, J.~E. 1990, \apj, 349, 562

\bibitem[{{Hernquist} \& {Ostriker}(1992)}]{HernquistOstriker1992}
{Hernquist}, L., \& {Ostriker}, J.~P. 1992, \apj, 386, 375

\bibitem[{{Kauffmann} \& {Haehnelt}(2000)}]{KauffmannHaehnelt2000}
{Kauffmann}, G., \& {Haehnelt}, M. 2000, \mnras, 311, 576

\bibitem[{{Magorrian} {et~al.}(1998){Magorrian}, {Tremaine}, {Richstone},
  {Bender}, {Bower}, {Dressler}, {Faber}, {Gebhardt}, {Green}, {Grillmair},
  {Kormendy}, \& {Lauer}}]{Magorrianetal1998}
{Magorrian}, J., {Tremaine}, S., {Richstone}, D., {Bender}, R., {Bower}, G.,
  {Dressler}, A., {Faber}, S.~M., {Gebhardt}, K., {Green}, R., {Grillmair}, C.,
  {Kormendy}, J., \& {Lauer}, T. 1998, \aj, 115, 2285

\bibitem[{{Makino}(1997)}]{Makino1997}
{Makino}, J. 1997, \apj, 478, 58

\bibitem[{{Makino} \& {Aarseth}(1992)}]{MakinoAarseth1992}
{Makino}, J., \& {Aarseth}, S.~J. 1992, \pasj, 44, 141

\bibitem[{{Makino} \& {Ebisuzaki}(1994)}]{MakinoEbisuzaki1994}
{Makino}, J., \& {Ebisuzaki}, T. 1994, \apj, 436, 607

\bibitem[{{Makino} \& {Ebisuzaki}(1996)}]{MakinoEbisuzaki1996}
---. 1996, \apj, 465, 527

\bibitem[{{Makino} {et~al.}(2002){Makino}, {Fukushige}, \&
  {Namura}}]{Makinoetal2002}
{Makino}, J., {Fukushige}, T., \& {Namura}, K. 2002, GRAPE-6: The
  massively-parallel special-purpose computer for astrophysical particle
  simulations, in preparation

\bibitem[{{Makino} {et~al.}(1993){Makino}, {Fukushige}, {Okumura}, \&
  {Ebisuzaki}}]{Makinoetal1993}
{Makino}, J., {Fukushige}, T., {Okumura}, S.~K., \& {Ebisuzaki}, T. 1993,
  \pasj, 45, 303

\bibitem[{{Makino} {et~al.}(1997){Makino}, {Taiji}, {Ebisuzaki}, \&
  {Sugimoto}}]{Makinoetal1997}
{Makino}, J., {Taiji}, M., {Ebisuzaki}, T., \& {Sugimoto}, D. 1997, \apj, 480,
  432

\bibitem[{{Mikkola} \& {Valtonen}(1992)}]{MikkolaValtonen1992}
{Mikkola}, S., \& {Valtonen}, M.~J. 1992, \mn, 259, 115

\bibitem[{{Milosavljevi{\' c}} \& {Merritt}(2001)}]{MilosavljevicMerritt2001}
{Milosavljevi{\' c}}, M., \& {Merritt}, D. 2001, \apj, 563, 34

\bibitem[{{Milosavljevi{\' c}} {et~al.}(2002){Milosavljevi{\' c}}, {Merritt},
  {Rest}, \& {van den Bosch}}]{Milosavljevicetal2002}
{Milosavljevi{\' c}}, M., {Merritt}, D., {Rest}, A., \& {van den Bosch}, F.~C.
  2002, \mnras, 331, L51

\bibitem[{{Nakano} \& {Makino}(1999{\natexlab{a}})}]{NakanoMakino1999a}
{Nakano}, T., \& {Makino}, J. 1999{\natexlab{a}}, \apj, 510, 155

\bibitem[{{Nakano} \& {Makino}(1999{\natexlab{b}})}]{NakanoMakino1999b}
---. 1999{\natexlab{b}}, \apjl, 525, L77



\bibitem[{{Okumura} {et~al.}(1991){Okumura}, {Ebisuzaki}, \&
  {Makino}}]{Okumuraetal1991}
{Okumura}, S.~K., {Ebisuzaki}, T., \& {Makino}, J. 1991, \pasj, 43, 781

\bibitem[{{Quinlan} \& {Hernquist}(1997)}]{QuinlanHernquist1997}
{Quinlan}, G.~D., \& {Hernquist}, L. 1997, New Astronomy, 2, 533

\bibitem[{{Saslaw} {et~al.}(1974){Saslaw}, {Valtonen}, \&
  {Aarseth}}]{SaslawValtonenAarseth1974}
{Saslaw}, W.~C., {Valtonen}, M.~J., \& {Aarseth}, S.~J. 1974, \apj, 190, 253

\bibitem[{Spitzer(1987)}]{Spitzer1987}
Spitzer, Lyman, J. 1987, Dynamical Evolution of Globular Clusters (Princeton,
  New Jersey: Princeton University Press)

\bibitem[{Springel {et~al.}(2001)Springel, Yoshida, \&
  White}]{Springeletal2000}
Springel, V., Yoshida, N., \& White, S.~D. 2001, New Astronomy, 6, 79

\bibitem[{{White}(1979)}]{White1979b}
{White}, S. D.~M. 1979, \mnras, 189, 831

\end{thebibliography}
\newcommand{\noopsort}[1]{} \newcommand{\printfirst}[2]{#1}
  \newcommand{\singleletter}[1]{#1} \newcommand{\switchargs}[2]{#2#1}

\end{document}